\documentclass[english]{article}
\usepackage[T1]{fontenc}
\usepackage[latin9]{inputenc}
\usepackage{geometry}
\usepackage{bm}
\usepackage{amsmath}
\usepackage{amsthm}
\usepackage{amssymb}
\usepackage{stmaryrd}
\usepackage[authoryear]{natbib}

\makeatletter

\providecommand{\tabularnewline}{\\}

\theoremstyle{plain}
\newtheorem{prop}{\protect\propositionname}

\usepackage{url}

\@ifundefined{showcaptionsetup}{}{%
 \PassOptionsToPackage{caption=false}{subfig}}
\usepackage{subfig}
\makeatother

\usepackage{babel}
\providecommand{\propositionname}{Proposition}

\begin{document}
\title{Universal inference with composite likelihoods}
\author{Hien D Nguyen\thanks{Corresponding author---Email: h.nguyen5@latrobe.edu.au. $^{1}$Department
of Mathematics and Statistics, La Trobe University, Melbourne, Australia.}$\text{ }^{1}$ \and Jessica Bagnall-Guerreiro$^{1}$ \and Andrew T Jones$^{1}$}
\maketitle
\begin{abstract}
Maximum composite likelihood estimation is a useful alternative to
maximum likelihood estimation when data arise from data generating
processes (DGPs) that do not admit tractable joint specification.
We demonstrate that generic composite likelihoods consisting of marginal
and conditional specifications permit the simple construction of composite
likelihood ratio-like statistics from which finite-sample valid confidence
sets and hypothesis tests can be constructed. These statistics are
universal in the sense that they can be constructed from any estimator
for the parameter of the underlying DGP. We demonstrate our methodology
via a simulation study using a pair of conditionally specified bivariate
models.
\end{abstract}

\section{Introduction}

Likelihood-based methods are among the most important tools for conducting
statistical inference. However, data generating processes (DGPs) of
complex models often do not admit tractable likelihood functions.
In such cases, a potential remedy is to specify the model based on
more amenable marginal and conditional probability density/mass functions
(PDFs/PMFs) of the DGP, instead. This joint specification is often
referred to as the composite likelihood (CL) or pseudolikelihood.

The literature regarding CL-based inference has its roots in the works
of \citet{Besag:1975aa} and \citet{Lindsay1988}. Further developments
regarding the theory and application of CL methods can be found in
\citet{Arnold1991}, \citet{Molenberghs2005}, \citet{Varin2011},
\citet{Yi:2014aa}, and \citet{Nguyen:2018ab}, among other works.

We build upon the recent work of \citet{Wasserman:2020aa} who demonstrated
the construction of sample splitting and sample swapping likelihood
ratio statistics that yield finite-sample valid confidence sets and
hypothesis tests, and are universal in the sense that they are agnostic
to parameter estimators. The inferential constructions are similar
to the recently popularized $e$-values of \citet{Vovk:2021wm}, as
well as the $s$-values of \citet{Grunwald:2020vf} and the betting
scores of \citet{Shafer:2021vh}. We demonstrate how our CL-based
methods can be used via applications to constructing confidence sets
and tests for a pair of conditionally specified bivariate models.
Here, we consider a simulation study regarding the exponential conditional
model of \citet{Arnold:1999aa} and the log-normal conditional model
of \citet{Sarabia:2007wu}.

The paper proceeds as follows. In Section 2, we present the CL framework
and the universal confidence set and hypothesis test constructions.
A simulation study of our methodology is presented in Section 3.

\section{Universal inference via composite likelihoods}

Let $\bm{X}\in\mathbb{X}\subseteq\mathbb{R}^{d}$ be a random variable
arising from a parametric distribution characterized by the PDF/PMF
(generically, PDF) $p\left(\bm{x};\bm{\theta}\right)$, where $\bm{\theta}\in\Theta\subseteq\mathbb{R}^{q}$
is a parameter vector ($d,q\in\mathbb{N}$). We shall write $\bm{X}^{\top}=\left(X_{1},\dots,X_{d}\right)$
to indicate a random variable and $\bm{x}^{\top}=\left(x_{1},\dots,x_{d}\right)$
to indicate its realization.

Let $2^{\left[d\right]}$ be the power set of $\left[d\right]=\left\{ 1,\dots,d\right\} $,
and let $\mathbb{S}_{d}=2^{\left[d\right]}\backslash\left\{ \varnothing\right\} $.
For each $\mathcal{S}\in\mathbb{S}_{d}$, let $\mathcal{S}=\left\{ s_{1},\dots,s_{\left|\mathcal{S}\right|}\right\} \subseteq\left[d\right]$,
where $\left|\mathcal{S}\right|$ is the cardinality of $\mathcal{S}$.
Further, let $\mathbb{T}_{d}$ be the set of all divisions of $\left[d\right]$
into two nonempty subsets. We represent each element of $\mathbb{T}_{d}$
as a pair $\mathcal{T}=\left(\overleftarrow{\mathcal{T}},\overrightarrow{\mathcal{T}}\right)$,
where $\overleftarrow{\mathcal{T}}=\left\{ \overleftarrow{t}_{1},\dots,\overleftarrow{t}_{\left|\overleftarrow{\mathcal{T}}\right|}\right\} \subset\left[d\right]$
and $\overrightarrow{\mathcal{T}}=\left\{ \overrightarrow{t}_{1},\dots,\overrightarrow{t}_{\left|\overrightarrow{\mathcal{T}}\right|}\right\} \subset\left[d\right]\backslash\overleftarrow{\mathcal{T}}$
are the 'left-hand' and 'right-hand' subsets of the division $\mathcal{T}$,
respectively. We note that $\left|\mathbb{S}_{d}\right|=2^{d}-1$
and $\left|\mathbb{T}_{d}\right|=3^{d}-2^{d+1}+1$.

For each $\mathcal{S}$ and $\mathcal{T}$, we assign a non-negative
coefficient $\sigma_{\mathcal{S}}$ and $\tau_{\mathcal{T}}$, respectively.
We call these coefficients the weights, and we put these weights into
the vectors $\bm{\sigma}=\left(\sigma_{\mathcal{S}}\right)_{\mathcal{S}\in\mathbb{S}_{d}}$
and $\bm{\tau}=\left(\tau_{\mathcal{T}}\right)_{\mathcal{T}\in\mathbb{T}_{d}}$,
respectively. We assume that $\upsilon=\sum_{\mathcal{S}\in\mathbb{S}_{d}}\sigma_{\mathcal{S}}+\sum_{\mathcal{T}\in\mathbb{T}_{d}}\tau_{\mathcal{T}}>0$.

Given weights $\bm{\sigma}$ and $\bm{\tau}$, we define the individual
CL (ICL) function for $\bm{X}$ as

\[
p_{\bm{\sigma},\bm{\tau}}\left(\bm{x};\bm{\theta}\right)=\prod_{\mathcal{S}\in\mathbb{S}_{d}}\left[p\left(\bm{x}_{\mathcal{S}};\bm{\theta}\right)\right]^{\sigma_{\mathcal{S}}/\upsilon}\prod_{\mathcal{T}\in\mathbb{T}_{d}}\left[p\left(\bm{x}_{\overleftarrow{\mathcal{T}}}|\bm{x}_{\overrightarrow{\mathcal{T}}};\bm{\theta}\right)\right]^{\tau_{\mathcal{T}}/\upsilon}\text{,}
\]
where $\bm{x}_{\mathcal{S}}^{\top}=\left(x_{s_{1}},\dots,x_{s_{\left|\mathcal{S}\right|}}\right)$,
$\bm{x}_{\overleftarrow{\mathcal{T}}}=\left(x_{\overleftarrow{t}_{1}},\dots,x_{\overleftarrow{t}_{\left|\overleftarrow{\mathcal{T}}\right|}}\right)$,
and $\bm{x}_{\overrightarrow{\mathcal{T}}}=\left(x_{\overrightarrow{t}_{1}},\dots,x_{\overrightarrow{t}_{\left|\overrightarrow{\mathcal{T}}\right|}}\right)$.
Here, $p\left(\bm{x}_{\mathcal{S}};\bm{\theta}\right)$ is the marginal
PDF of $\bm{X}_{\mathcal{S}}$, and $p\left(\bm{x}_{\overleftarrow{\mathcal{T}}}|\bm{x}_{\overrightarrow{\mathcal{T}}};\bm{\theta}\right)$
is the conditional PDF of $\bm{X}_{\overleftarrow{\mathcal{T}}}$
conditioned on $\bm{X}_{\overrightarrow{\mathcal{T}}}=\bm{x}_{\overrightarrow{\mathcal{T}}}$.

\subsection{Sample splitting and sample swapping}

Let $\mathbf{X}_{n}=\left(\bm{X}_{i}\right)_{i=1}^{n}$ be a sequence
of $n$ IID random variables with the same DGP as $\bm{X}$, and split
$\mathbf{X}_{n}$ into two subsamples $\mathbf{X}_{n}^{1}=\left(\bm{X}_{i}^{1}\right)_{i=1}^{n_{1}}$
and $\mathbf{X}_{n}^{2}=\left(\bm{X}_{i}^{2}\right)_{i=1}^{n_{2}}$
of sizes $n_{1}$ and $n_{2}$, respectively, where $n=n_{1}+n_{2}$.
We assume that $\bm{X}$ has a DGP that is characterized by the PDF
$p\left(\bm{x};\bm{\theta}_{0}\right)$, for some $\bm{\theta}_{0}\in\Theta$,
and we let $\Pr_{\bm{\theta}_{0}}$ be its corresponding probability
measure. Let $\hat{\bm{\theta}}_{n}^{1}$ and $\hat{\bm{\theta}}_{n}^{2}$
be a pair of generic estimators of $\bm{\theta}_{0}$, using only
$\mathbf{X}_{n}^{1}$ or $\mathbf{X}_{n}^{2}$, respectively.

For $k\in\left\{ 1,2\right\} $, we let
\[
L_{\bm{\sigma},\bm{\tau}}\left(\bm{\theta};\mathbf{X}_{n}^{k}\right)=\prod_{i=1}^{n_{k}}p_{\bm{\sigma},\bm{\tau}}\left(\bm{X}_{i}^{k}\right)
\]
be the CL function of $\mathbf{X}_{n}^{k}$, as a function of $\bm{\theta}$.
We write the split sample CL ratio statistics (spCLRSs) and the swapped
sample CL ratio statistic (swCLRS) as 
\[
U_{\bm{\sigma},\bm{\tau}}^{k}\left(\bm{\theta};\mathbf{X}_{n}\right)=L_{\bm{\sigma},\bm{\tau}}\left(\hat{\bm{\theta}}^{3-k};\mathbf{X}_{n}^{k}\right)/L_{\bm{\sigma},\bm{\tau}}\left(\bm{\theta};\mathbf{X}_{n}^{k}\right)\text{,}
\]
for each $k\in\left\{ 1,2\right\} $, and 
\[
\bar{U}_{\bm{\sigma},\bm{\tau}}\left(\bm{\theta};\mathbf{X}_{n}\right)=\left\{ U_{\bm{\sigma},\bm{\tau}}^{1}\left(\bm{\theta};\mathbf{X}_{n}\right)+U_{\bm{\sigma},\bm{\tau}}^{2}\left(\bm{\theta};\mathbf{X}_{n}\right)\right\} /2\text{,}
\]
respectively.

For $\alpha\in\left(0,1\right)$, let
\[
\mathcal{C}^{\alpha}\left(\mathbf{X}_{n}\right)=\left\{ \bm{\theta}\in\Theta:U_{\bm{\sigma},\bm{\tau}}^{1}\left(\bm{\theta};\mathbf{X}_{n}\right)\le1/\alpha\right\} \text{ and }\bar{\mathcal{C}}^{\alpha}\left(\mathbf{X}_{n}\right)=\left\{ \bm{\theta}\in\Theta:\bar{U}_{\bm{\sigma},\bm{\tau}}\left(\bm{\theta};\mathbf{X}_{n}\right)\le1/\alpha\right\} 
\]
be confidence sets based on the spCLRS and the swCLRS, respectively.
We have the following result regarding the validity of $\mathcal{C}^{\alpha}\left(\mathbf{X}_{n}\right)$
and $\bar{\mathcal{C}}^{\alpha}\left(\mathbf{X}_{n}\right)$ (all
theoretical results in this work are proved in \citealp{Nguyen:2020uk}).
\begin{prop}
\label{prop: confidence}The set estimators $\mathcal{C}^{\alpha}\left(\mathbf{X}_{n}\right)$
and $\bar{\mathcal{C}}^{\alpha}\left(\mathbf{X}_{n}\right)$ are finite
sample-valid $100\left(1-\alpha\right)\%$ confidence sets for $\bm{\theta}_{0}$
in the sense that
\[
\mathrm{Pr}_{\bm{\theta}_{0}}\left(\bm{\theta}_{0}\in\mathcal{C}^{\alpha}\left(\mathbf{X}_{n}\right)\right)\ge1-\alpha\text{, and }\mathrm{Pr}_{\bm{\theta}_{0}}\left(\bm{\theta}_{0}\in\mathcal{\bar{C}}^{\alpha}\left(\mathbf{X}_{n}\right)\right)\ge1-\alpha\text{,}
\]
for any $n\in\mathbb{N}$.
\end{prop}
We now consider the testing of null and alternative hypotheses
\[
\text{H}_{0}:\bm{\theta}\in\Theta_{0}\text{ and }\text{H}_{1}:\bm{\theta}\in\Theta_{1}\text{,}
\]
where $\Theta_{0},\Theta_{1}\subseteq\Theta$. Let
\[
\mathbb{M}\left(\mathbf{X}_{n}^{k}\right)=\left\{ \bm{\theta}\in\Theta_{0}:L_{\bm{\sigma},\bm{\tau}}\left(\bm{\theta};\mathbf{X}_{n}^{k}\right)=\max_{\bm{\vartheta}\in\Theta_{0}}L_{\bm{\sigma},\bm{\tau}}\left(\bm{\vartheta};\mathbf{X}_{n}^{k}\right)\right\} 
\]
be the set of maximizers of the CL function $L_{\bm{\sigma},\bm{\tau}}\left(\bm{\theta};\mathbf{X}_{n}^{k}\right)$,
for each $k\in\left\{ 1,2\right\} $, and write $\tilde{\bm{\theta}}_{n}^{k}\in\mathbb{M}\left(\mathbf{X}_{n}^{k}\right).$
We then write the sample splitting and sample swapping test statistics
as
\[
V_{\bm{\sigma},\bm{\tau}}^{k}\left(\mathbf{X}_{n}\right)=U_{\bm{\sigma},\bm{\tau}}^{k}\left(\tilde{\bm{\theta}}_{n}^{k}\right)\text{, and }\bar{V}_{\bm{\sigma},\bm{\tau}}\left(\mathbf{X}_{n}\right)=\left\{ U_{\bm{\sigma},\bm{\tau}}^{1}\left(\tilde{\bm{\theta}}_{n}^{1}\right)+U_{\bm{\sigma},\bm{\tau}}^{2}\left(\tilde{\bm{\theta}}_{n}^{2}\right)\right\} /2\text{,}
\]
respectively. Further, define the split sample CL ratio test (spCLRT)
and the swapped sample CL ratio test (swCLRT) by the rejection rules:
reject $\text{H}_{0}$ if $V_{\bm{\sigma},\bm{\tau}}^{1}\left(\mathbf{X}_{n}\right)\ge1/\alpha$
or if $\bar{V}_{\bm{\sigma},\bm{\tau}}\left(\mathbf{X}_{n}\right)\ge1/\alpha$,
respectively. We have the following result regarding the finite sample-validity
of the tests.
\begin{prop}
\label{prop: test}The spCLRT and swCLRT control the Type I error
for all $\alpha\in\left(0,1\right)$ and $n\in\mathbb{N}$ in the
sense that
\[
\sup_{\bm{\theta}_{0}\in\Theta_{0}}\mathrm{Pr}_{\bm{\theta}_{0}}\left(V_{\bm{\sigma},\bm{\tau}}^{1}\left(\mathbf{X}_{n}\right)>1/\alpha\right)\le\alpha\text{, and }\sup_{\bm{\theta}_{0}\in\Theta_{0}}\mathrm{Pr}_{\bm{\theta}_{0}}\left(\bar{V}_{\bm{\sigma},\bm{\tau}}\left(\mathbf{X}_{n}\right)>1/\alpha\right)\le\alpha\text{.}
\]
\end{prop}

\section{Simulation study}

All numerical computation were conducted in the $\mathsf{R}$ programming
environment \citep{R-Core-Team:2020aa}. The code for the analyses
are made available at \url{hiendn / CompositeLikelihoodISI}.

\subsection{Bivariate distribution with exponential conditional distributions}

We first consider a simulation study regarding data generated from
the bivariate exponential distribution of \citet[Sec. 4.4]{Arnold:1999aa}.
Here the random variable $\bm{X}^{\top}=\left(X_{1},X_{2}\right)$
has joint PDF
\[
p\left(\bm{x};\theta\right)=\kappa\left(\theta\right)\exp\left\{ -x_{1}-x_{2}-\theta x_{1}x_{2}\right\} \text{,}
\]
where $\theta\ge0$ is the parameter of interest, and $\kappa\left(\theta\right)=\theta\exp\left\{ -1/\theta\right\} /\int_{1/\theta}^{\infty}w^{-1}\exp\left(-w\right)\text{d}w$
is an intractable normalization constant. However, the conditional
PDFs of $X_{k}|X_{3-k}=x_{3-k}$, for $k\in\left\{ 1,2\right\} $,
can be specified by
\[
p\left(x_{k}|x_{3-k};\theta\right)=f_{\text{Exp}}\left(x_{k};1+\theta x_{3-k}\right)\text{,}
\]
where $f_{\text{Exp}}\left(x;\lambda\right)=\lambda\exp\left(-\lambda x\right)$
is the PDF of the exponential distribution with rate $\lambda>0$.
Thus, we can conduct inference regarding this DGP by considering ICLs
of the form

\[
p_{\bm{\sigma},\bm{\tau}}\left(\bm{x};\theta\right)=\left[p\left(x_{1}|x_{2};\theta\right)\right]^{1/2}\left[p\left(x_{2}|x_{1};\theta\right)\right]^{1/2}\text{,}
\]
where $\bm{\sigma}=\mathbf{0}$ and $\bm{\tau}=\left(1/2\right)\mathbf{1}$.

For data $\mathbf{X}_{n}$ with identical DGP to $\bm{X}$, characterized
by $\theta_{0}\in\left\{ 1,5,10\right\} $, where $n_{1}=n_{2}\in\left\{ 100,1000,10000\right\} $,
we consider the use of the spCLRS and swCLRS confidence sets at the
$\alpha=0.05$ level. Here, each confidence set is constructed using
the maximum composite likelihood estimator (MCLE).

For each pair $\left(n_{1},\theta\right)$, we replicate the simulation
$r=100$ times and compute the coverage proportion (CP) and average
size (AS) of the confidence intervals for the two set constructions.
Here, CP and AS are computed as $r^{-1}\sum_{j=1}^{r}\left\llbracket \theta_{0}\in\mathcal{C}_{j}\right\rrbracket $
and $r^{-1}\sum_{j=1}^{r}\text{diam}\left(\mathcal{C}_{j}\right)$,
where $\mathcal{C}_{j}$ is a stand-in for a confidence set constructed
from the $r\text{th}$ replicate, $\left\llbracket \cdot\right\rrbracket $
are Iverson brackets, and $\text{diam}\left(\cdot\right)$ is the
metric set diameter operator.

The results are presented in Table \ref{tab:CP-and-AS}(a). We observe
that CP was near perfect, with only one scenario yielding a confidence
set that did not contain $\theta_{0}$. This supports Proposition
\ref{prop: confidence}, although it indicates that the confidence
sets are fairly conservative. We observe that AS is decreasing in
$n_{1}$, as expected, and increasing in $\theta_{0}$. We also find
that the swCLRS sets are smaller than the spCLRS sets, which suggests
a more efficient use of the data.

\begin{table}
\caption{\label{tab:CP-and-AS}Simulation results.}

\subfloat[CP and AS results for the spCLRS and swCLRS $95\%$ confidence sets.]{
\begin{centering}
\begin{tabular}{|cc||ccc|ccc|}
\hline 
 &  & CP &  & $n_{1}$ & AS &  & $n_{1}$\tabularnewline
 & $\theta_{0}$ & 100 & 1000 & 10000 & 100 & 1000 & 10000\tabularnewline
\hline 
\hline 
spCLRS & 1 & 1 & 1 & 1 & 1.43 & 0.46 & 0.14\tabularnewline
 & 5 & 1 & 1 & 1 & 4.60 & 1.49 & 0.47\tabularnewline
 & 10 & 1 & 1 & 1 & 8.32 & 2.57 & 0.82\tabularnewline
\hline 
swCLRS & 1 & 1 & 1 & 1 & 1.28 & 0.40 & 0.12\tabularnewline
 & 5 & 1 & 0.99 & 1 & 4.13 & 1.29 & 0.40\tabularnewline
 & 10 & 1 & 1 & 1 & 7.40 & 2.31 & 0.73\tabularnewline
\hline 
\end{tabular}
\par\end{centering}
}\subfloat[Proportion of rejections by the spCLRT and swCLRT.]{
\centering{}%
\begin{tabular}{|cc||ccc|}
\hline 
 &  & Rej. &  & $n_{1}$\tabularnewline
 & $c_{0}$ & 100 & 1000 & 10000\tabularnewline
\hline 
\hline 
spCLRT & 0 & 0 & 0 & 0\tabularnewline
 & 1 & 0.26 & 1 & 1\tabularnewline
 & 5 & 0.98 & 1 & 1\tabularnewline
\hline 
swCLRT & 0 & 0 & 0 & 0\tabularnewline
 & 1 & 0.32 & 1 & 1\tabularnewline
 & 5 & 1 & 1 & 1\tabularnewline
\hline 
\end{tabular}}

\end{table}

\subsection{Bivariate distribution with log-normal conditional distributions}

We now consider the bivariate distribution of \citet{Sarabia:2007wu},
which is specified by the PDF
\begin{equation}
p\left(\bm{x};\bm{\theta}\right)=\frac{\kappa\left(c\right)}{2\pi\sigma_{1}\sigma_{2}x_{1}x_{2}}\exp\left\{ -\frac{1}{2}\left[\left(\frac{\log x_{1}-\mu_{1}}{\sigma_{1}}\right)^{2}+\left(\frac{\log x_{2}-\mu_{2}}{\sigma_{2}}\right)^{2}+c\left(\frac{\log x_{1}-\mu_{1}}{\sigma_{1}}\right)^{2}\left(\frac{\log x_{2}-\mu_{2}}{\sigma_{2}}\right)^{2}\right]\right\} \text{,}\label{eq: lognormal joint}
\end{equation}
where $\bm{\theta}^{\top}=\left(\mu_{1},\sigma_{1}^{2},\mu_{2},\sigma_{2}^{2},c\right)$,
with $\mu_{1},\mu_{2}\in\mathbb{R}$, $\sigma_{1}^{2},\sigma_{2}^{2}>0$,
and $c\ge0$. Here, $\kappa\left(c\right)=\sqrt{2c}/U\left(1/2,1,\left(2c\right)^{-1}\right)$,
where $U\left(a,b,z\right)$ is the confluence hypergeometric function,
defined as per \citet[Eqn. 13.2.5]{AbramowitzStegun:1972vl}. Like
in the previous example, the normalizing constant of the joint PDF
makes it intractable. However, we may again specify the conditional
PDFs of $X_{k}|X_{3-k}=x_{3-k}$, for $k\in\left\{ 1,2\right\} $,
by
\[
p\left(x_{k}|x_{3-k};\bm{\theta}\right)=f_{\text{LN}}\left(x_{k};\mu_{k},\sigma_{k}^{2}/\left\{ 1+c\left(\frac{\log x_{3-k}-\mu_{3-k}}{\sigma_{3-k}}\right)^{2}\right\} \right)\text{,}
\]
where
\[
f_{\text{LN}}\left(x;\mu,\sigma^{2}\right)=\frac{1}{x\sqrt{2\pi\sigma^{2}}}\exp\left\{ -\frac{1}{2}\left(\frac{\log x-\mu}{\sigma}\right)^{2}\right\} 
\]
is the PDF of a log-normal distribution with location and scale parameters
$\mu\in\mathbb{R}$ and $\sigma^{2}>0$, respectively. We can use
the conditional PDFs to conduct CL inference via the ICLs of the form
\[
p_{\bm{\sigma},\bm{\tau}}\left(\bm{x};\bm{\theta}\right)=\left[p\left(x_{1}|x_{2};\bm{\theta}\right)\right]^{1/2}\left[p\left(x_{2}|x_{1};\bm{\theta}\right)\right]^{1/2}\text{,}
\]
where $\bm{\sigma}=\mathbf{0}$ and $\bm{\tau}=\left(1/2\right)\mathbf{1}$. 

We simulate data $\mathbf{X}_{n}$, $n_{1}=n_{2}\in\left\{ 100,1000,10000\right\} $
from DGPs that are characterized by the PDF (\ref{eq: lognormal joint}),
with parameter vector $\bm{\theta}_{0}=\left(2,1,2,1,c_{0}\right)$,
where $c_{0}\in\left\{ 0,1,5\right\} $. For each pair $\left(n_{1},c_{0}\right)$,
we use the spCLRT and swCLRT to test the hypotheses $\text{H}_{0}:c_{0}=0$
versus $\text{H}_{1}:c_{0}>0$, at the $\alpha=0.05$ level. We repeat
each simulation pair $r=100$ times and compute the proportion of
times the null hypothesis was rejected. Here, we again make use of
the MCLE. 

The results are reported in Table \ref{tab:CP-and-AS}(b). Notice
that no false rejections were made when $c_{0}=0$, thus the size
of the test is conservatively controlled, as predicted by Proposition
\ref{prop: test}. We also see that the tests become increasingly
powerful as $c_{0}$ increases and as $n_{1}$ increases, as would
be expected. There is some evidence that the swCLRT is more powerful
than the spCLRT, conforming to observations from the previous study.

\bibliographystyle{plainnat}
\bibliography{20201215_MASTERBIB}

\end{document}